\documentclass[useAMS]{mn2e}

\def\Msol{M$_\odot$\ }
\def\mnras{MNRAS}
\def\apj{ApJ}
\def\apjl{ApJL}

\def\aap{A\& A}
\def\aapr{A\& A Rev}

\def\pasj{PASJ}
\def\nat{Nature}
\usepackage{times}
\usepackage{amssymb}
\usepackage{rotating}
\usepackage{amsmath}
\input{epsf}

\title[Thermal Winds in Binaries]
{Thermal Winds in Stellar Mass Black Hole and Neutron Star Binary Systems}

\author[C. Done, R. Tomaru, T. Takahashi]
{Chris Done$\thanks{E-mail:chris.done@durham.ac.uk}^{1,2}$,
Ryota Tomaru$^{2,3}$ \&
Tadayuki Takahashi$^2$
\\
$^1$ Centre for Extragalactic Astronomy, 
Department of Physics, University of Durham, South Road,
Durham DH1 3LE, UK\\
$^2$ Institute of Space and Astronautical Science (ISAS), Japan Aerospace Exploration Agency (JAXA), Kanagawa 252-5210, Japan\\
$^3$ Department of Physics, University of Tokyo, 7-3-1 Hongo, Bunkyo, Tokyo, 113-0033, Japan.\\
}

\date{Submitted to MNRAS}
\pagerange{\pageref{firstpage}--\pageref{lastpage}} \pubyear{2016}

\begin{document}

\topmargin = -0.5cm

\maketitle

\label{firstpage}

\begin{abstract}
Black hole binaries show equatorial disc winds at high
luminosities, which apparently disappear during the spectral
transition to the low/hard state. This is also where the radio jet
appears, motivating speculation that both wind and jet are driven by
different configurations of the same magnetic field.  However, these
systems must also have thermal winds, as the outer disc is clearly
irradiated. We develop a predictive model of the absorption features
from thermal winds, based on pioneering work of Begelman et al
1983. We couple this to a realistic model of the irradiating spectrum
as a function of luminosity to predict the entire wind evolution
during outbursts. We show that the column density of the thermal wind
scales roughly with luminosity, and does not shut off at the spectral
transition, though its visibility will be affected by the abrupt
change in ionising spectrum. We re-analyse the data from H1743-322
which most constrains the difference in wind across the spectral
transition and show that these are consistent with the thermal wind
models. We include simple corrections for radiation pressure, which
allows stronger winds to be launched from smaller radii. These winds
become optically thick around Eddington, which may even explain the
exceptional wind seen in one observation of GRO J1655-40.
These data can instead be fit by magnetic wind models, but similar
winds are not seen in this or other systems at
similar luminosities. Hence we conclude that the majority (perhaps all)
current data can be explained by thermal or thermal-radiative winds.

\end{abstract}


%==============================================
\section{Introduction} \label{sec:introduction}
%==============================================

Absorption lines from ionised material are seen in many high
inclination Low Mass X-ray Binary systems, both neutron stars (NS) and
black holes (BHB). These are most evident in the deep absorption dips
which occur at specific orbital phases due to clumps formed where the
accretion stream hits the disc.  However, some absorption lines remain
even outside of the dip events. These are seen mainly as hydrogen and
helium-like iron K$\alpha$, indicating very highly ionised material
which is
distributed fairly uniformly around all azimuths. These lines can be
blue shifted by up to a few thousand km/s, showing clearly that there
is an equatorial disc wind in these systems, where the material is
strongly photo-ionised by the X-ray illumination from the central
source (see e.g. the reviews by Diaz Trigo \& Boirin 2013; 2016 and
Ponti et al 2012).

The most important questions about these winds concerns how they are
launched, and their connection to the accretion flow and its jet. Both
black hole and neutron star binaries show a dramatic spectral
transition, from a disc dominated spectrum at high luminosity to a
much harder Comptonised spectrum at lower luminosities. This can be
explained as the transition between a disc and hot accretion flow (see
e.g. the review by Done, Gierlinski \& Kubota 2007).  Data from
GRS1915+105 was first used to study the change in wind properties with
spectral state, with Neilson \& Lee (2009) showing that the wind
disappeared as the source made a transition to the harder spectral
state. They noted that this spectral transition also marks the onset
of the compact radio jet, so speculated on the existence of a causal
link, with the change in accretion flow properties also changing the
magnetic field configuration so that the same magnetically driven 
outflow is either a wind
in the soft, disc dominated state or a jet in the low/hard state.
Miller et al (2012) similarly find that the wind is not present in a
hard state of H1743-322, while it is clearly evident in a soft
state from the same source. The systematic survey of existing data by
Ponti et al (2012) showed that none of the sources with winds in the
high/soft state had significant absorption 
features in the low/hard state, though there are
not many sources, and the most constraining data are from GRS1915+105
and H1743-322 discussed above. 

An alternative explanation for the lack of wind features in the low/hard state 
is that the changing spectral shape at the transition
changes the ionisation of the wind (e.g. Diaz Trigo
et al. 2014). However, other photoionisation calculations show that
the difference in wind properties between hard and soft spectral state
cannot be completely explained by the same wind being present at
different ionisation level (Chakravorty et al 2013; Higginbottom \&
Proga 2015). This motivated studies of magnetic winds (Chakravorty et
al 2016), though it is now clear that both winds and jets can co-exist
at high luminosities (Kalemci et al 2016, Homan et al 2016).

However, winds in binary systems need not be magnetically
driven. X-ray irradiation of the outer disk can produce a
thermally driven wind/corona. The X-ray flux from the innermost
regions illuminates the upper layers of the outer disk, heating it up
to the Compton temperature, $T_{\rm IC}$, which depends only on
spectral shape. The heated upper layer expands on the sound speed
$c_{IC}=(kT_{IC}/\mu)^{1/2}$ where $\mu$ is the mean particle mass
(ions and electrons), fixed at $0.61m_p$ for solar abundances. The
material is then unbound at radii where this sound speed 
is larger than the escape speed i.e. $R>R_{IC}=
GM/c_{IC}^2$. This gives $R_{IC}=6.4\times 10^4/T_{IC,8} R_g$ (where $R_g=GM/c^2$ and
$T_{IC,8}=T_{IC}/10^8~{\rm K}$) so the material escapes as a wind
(Begelman et al.\ 1983, hereafter B83). A more careful treatment shows that the wind
can be launched from $R>0.2R_{IC}$ as long as the luminosity is high
enough to sustain rapid heating (B83, Woods et al
1996, hereafter W96).  Conversely, at smaller radii/lower luminosity the 
majority of the material
remains bound but forms an extended atmosphere above the disk
(B83, W96; Jimenez-Garate et al 2002). 

Current data from the high inclination NS systems show a good
qualitative match to the thermal wind predictions, with small disc
systems (short period binaries) having absorbing material which is
static, while outflows are only seen at radii larger than $0.2R_{IC}$
(Diaz Trigo \& Boirin 2013; 2016). Many winds in the black hole
systems also have a fairly large launch radius, consistent with
thermal driving (e.g. Kubota et al 2007; Diaz Trigo \& Boirin 2016),
and the most powerful winds are seen in the systems with the biggest
discs (Diaz Trigo \& Boirin 2016). This is consistent with the
predictions of thermal winds (see below), but not a natural consequence of
magnetic wind models.

However, there is a single observation of dramatic absorption from GRO
J1655-40 at low luminosities.  The material has high column, and lower
ionisation state than is normally seen, so has multiple lines from
lower atomic number elements as well as the standard H- and He-like
iron absorption. These require that the wind is launched from $<<
0.1R_{IC}$ if the observed luminosity is a good estimate of the
intrinsic luminosity, so this cannot be a thermal wind (Miller et al
2006; Kallman et al 2009; Luketic et al 2010; Neilsen \& Homan 2012;
Higginbottom \& Proga 2015).  However, the unusual properties of the
broadband continuum seen during this observation indicate that the
intrinsic source luminosity may be substantially underestimated (Done,
Gierlinski \& Kubota 2007; Shidatsu et al 2016; Neilsen et al
2016). If the intrinsic source luminosity is actually close to
Eddington then strong winds can be launched from much smaller radii
than predicted by the thermal wind models. Electron scattering in a
completely ionised, optically thick inner wind could suppress the
observed luminosity to below Eddington, with the observed absorption
lines tracing only lower ionisation material in the outer photosphere.
Whatever the origin of the wind in this single observation, it does
not represent the majority of winds seen in binaries, nor is it seen
in observations at similar luminosity of the same source (Nielsen \&
Homan 2012). Yet this single dataset, coupled with the observed
anti-correlation between wind and jet discussed above, has led to a
focus on magnetic winds in the current literature.

Even if there are magnetic winds, thermal winds should also be present
(e.g. Neilsen \& Homan 2012), as we know that the outer disc is
illuminated - the observed optical emission is dominated by
reprocessed X-ray flux (van Paradijs \& McClintock 1994).  In marked
contrast with magnetic winds, thermal winds are rather well understood
theoretically in terms of overall mass loss rates. However, models for
thermal winds were developed long before the advent of detectors which
could observe them, and they have been mostly sidelined due to the
focus on magnetic winds, with only a few recent papers on their
properties (Luketic et al 2010; Higginbottom \& Proga 2015). 

Here, we combine analytic and numerical thermal wind models to make
quantitative predictions for the mass loss rates for thermal winds,
and their most basic observables (i.e. column density and ionisation state).  We
couple this to a simple model of how the spectrum changes with
luminosity during outbursts to quantitatively explore the effect of
the spectral transition. The thermal wind models roughly predict that
the column density of the wind is proportional to the mass accretion
rate.  Thus the amount of material in the wind should not change much
as the source switches from high/soft to low/hard at roughly constant
luminosity during the spectral tranistion, though its visibility will
be affected by the change in photo-ionising spectrum (Chakravorty et
al 2013; Diaz Trigo et al 2014; Higginbottom \& Proga 2015). We
compare this to the current observational constraints from H1743-322
on the wind in a high/soft state being supressed in a low/hard state
(Miller et al 2012). We find that the thermal wind models are
consistent with their data as this high/soft state is probably an
order of magnitude higher mass accretion rate than the comparison
low/hard state, so should have an order of magnitude stronger wind. A
more stringent test would be to follow the wind evolution during a
single transition, rather than to compare the wind seen in low/hard
and high/soft spectra at different mass accretion rates separated by
years.

We incorporate a simple correction for radiation pressure as the
source approaches the Eddington limit. This allows the wind to be
launched from smaller radii than $R<0.2R_{IC}$ as it reduces the
effective gravity.  We show that these thermal-radiative winds do
indeed become optically thick as $L\to L_{Edd}$, so they could
potentially explain the anomolous wind in GROJ1655-40.  GRS1915+105 is
similarly close to Eddington, but its more complex and fast variable
spectra require a specialised analysis which is beyond the scope of
this paper (e.g. Ueda et al 2010; Zoghbi et al 2016). These are the
three main systems claimed to rule out the thermal wind models and
hence require magnetic driving.  We conclude that this is premature,
and that thermal (and thermal-radiative) wind models already explain
the vast majority (and perhaps all) of winds currently seen in binary systems.

%==============================================
\section{Thermal wind models}
%==============================================

B83 discuss the different physical regimes in which
Compton heated, thermal winds can form.  We repeat here their analysis
for completeness, pulling out only the required terms from their much
longer, more detailed analysis.

The central source spectrum contains photons over a broad range of
energies, and illuminates the disc at distance $R$ with differential
photon number rate $N(\nu)$ corresponding to energy flux $F(\nu)=h\nu
N(\nu)=L(\nu)/(4\pi R^2)$ in ergs~cm$^2$~s$^{-1}$~Hz$^{-1}$ or
$L(E)/(4\pi R^2)$ in ergs~cm$^2$~s$^{-1}$~erg$^{-1}$.  Electrons in
the disc photosphere at this radius have initial temperature $T_e$.
Photons with energies $E=h\nu\gg T_e$ will Compton downscattered,
losing some energy to the disc, typically $\Delta
E/E=-E/(m_ec^2)$. Conversely, those at low energies will be Compton
upscattered by the hot electrons in the disc, with typical $\Delta
E/E=4 kT_e/(m_ec^2)$, so they cool the photosphere. Combing these
gives an approximation for the energy change from both processes as
$\Delta E=4 kT_e E/(m_ec^2) -E^2/(m_ec^2)$. This reaches steady state
when heating balances cooling, so $\int \Delta E N(E) dE=0$, defining
the Compton temperature $kT_{IC}=\frac{1}{4}\int EL(E) dE/L$ where
$L=\int L(E)dE$ (see e.g. Done 2010).

Compton cooling is the dominant cooling process only at high
temperatures, where the material is completely ionised.  Deeper down
in the photosphere the heating from irradiation will be lower as the
upper layers have already absorbed some of the energy, so the gas temperature, $T_e$,
goes down.  The disc is in
hydrostatic equilibrium so the gas pressure, $P_{gas}=nkT_e$, must be higher to support the
weight of the 
upper layers. Thus the density, $n$, must increase downwards by more than the decrease in 
temperature from the lower irradiation. Higher density
means that Bremsstrahlung cooling becomes important, which lowers the
temperature still further, requiring an even larger increase in density to give
the required pressure support for the weight of the upper
layers. Eventually the temperature is so low that electrons can start
to be bound to ions, making line cooling possible. This triggers an
ionisation instability, as the line cooling lowers the
temperature, but this allows more electrons to be bound, making more
line transitions and hence more cooling (Krolik, McKee \& Tarter 1981). Thus the disc
photosphere splits into a hot, high ionisation skin overlying a cool,
dense photosphere (Nayakshin, Kazanas \& Kallman 2000). The typical
temperature of the skin is the Compton temperature $kT_{IC}$, and its typical
pressure can be calculated from the pressure ionisation parameter
$\Xi=P_{rad}/P_{gas} = L/(4\pi R^2 c P_{gas})$. The instability is
triggered at $\Xi\sim 40$, giving the gas density $n=L/(160\pi R^2 c kT_{IC})$ (B83;
W96).

The mass loss per unit area, $\dot{m}$ is then driven by the material
in the skin expanding on the thermal sound speed. For an isothermal
flow, the pressure at the sonic point is a factor of 2 lower than at
the base, so this gives $\dot{m}=\frac{1}{2}n \mu c_{IC}$. The total
mass loss rate in the wind is then

\begin{equation}
\begin{split}
\dot{M}_{wind}=\int_{R_{in}}^{R_{out}} \dot{m} \times 2\times 2\pi R dR\\
\approx  \frac{L/c}{80\ c_{IC}}\log R_{out}/R_{in}
\end{split}
\label{eqn:mdota}
\end{equation}
where the factor of $2$ comes from the fact that the disc has two
sides. From the discussion above, $R_{in}\sim 0.2R_{IC}$ and $R_{out}$ is the disc outer radius, 
and the total mass loss rate in the wind is directly proportional to the 
source luminosity. 

However, the wind is only isothermal if it is heated sufficiently
quickly. This depends on the irradiating flux, which drops with
increasing radius, so the wind is not heated so efficiently. The
Compton heating rate on each electron, $\Gamma_e$, is the incident
photon flux, $\sigma_T L/(\langle E \rangle 4\pi R^2)$, where $\langle
E\rangle$ is the mean photon energy, times the mean increase in energy
for each photon collision $\approx 4kT_{IC} \langle E\rangle
/(m_ec^2)$ from the Compton heating discussion above. Hence
$\Gamma_e=\sigma_T L kT_{IC} /(\pi R^2 m_ec^2)$.  For large
luminosities, the material is heated impulsively and reaches the
Compton temperature at the isothermal sonic point which is close to
the disc, as assumed above. For lower luminosities, electrons in the
gas are heated steadily, reaching a characteristic energy
$kT_{ch}=\Gamma_e R/c_{ch}$ where $R/c_{ch}$ is the time taken for the
material to reach height $\sim R$. This equation determines the
characteristic temperature, $T_{ch}$ and its corresponding sound speed
$c_{ch}=(kT_{ch}/\mu)^{1/2}$, but the gas pressure above the
ionisation instability is still set by the radiation flux as before so
$P_{ch}=nkT_{ch}= L/(160\pi R^2 c)$. This gives
\begin{equation}
kT_{ch} =kT_{IC}
(L/L_{crit})^{2/3} \zeta^{-2/3}
\end{equation} 
where $L_{crit}$ is the luminosity which
is just able to heat the gas to $kT_{IC}$ as it reaches height $\sim
R$ so that it is able to escape, at distance $\zeta=R/R_{IC}=1$.  
Equivalently, this gives
\begin{equation}
c_{ch} =c_{IC}
(L/L_{crit})^{1/3} \zeta^{-1/3}
\label{eqn:cch}
\end{equation}

\begin{figure}
\begin{tabular}{c}
\epsfxsize=0.45\textwidth \epsfbox{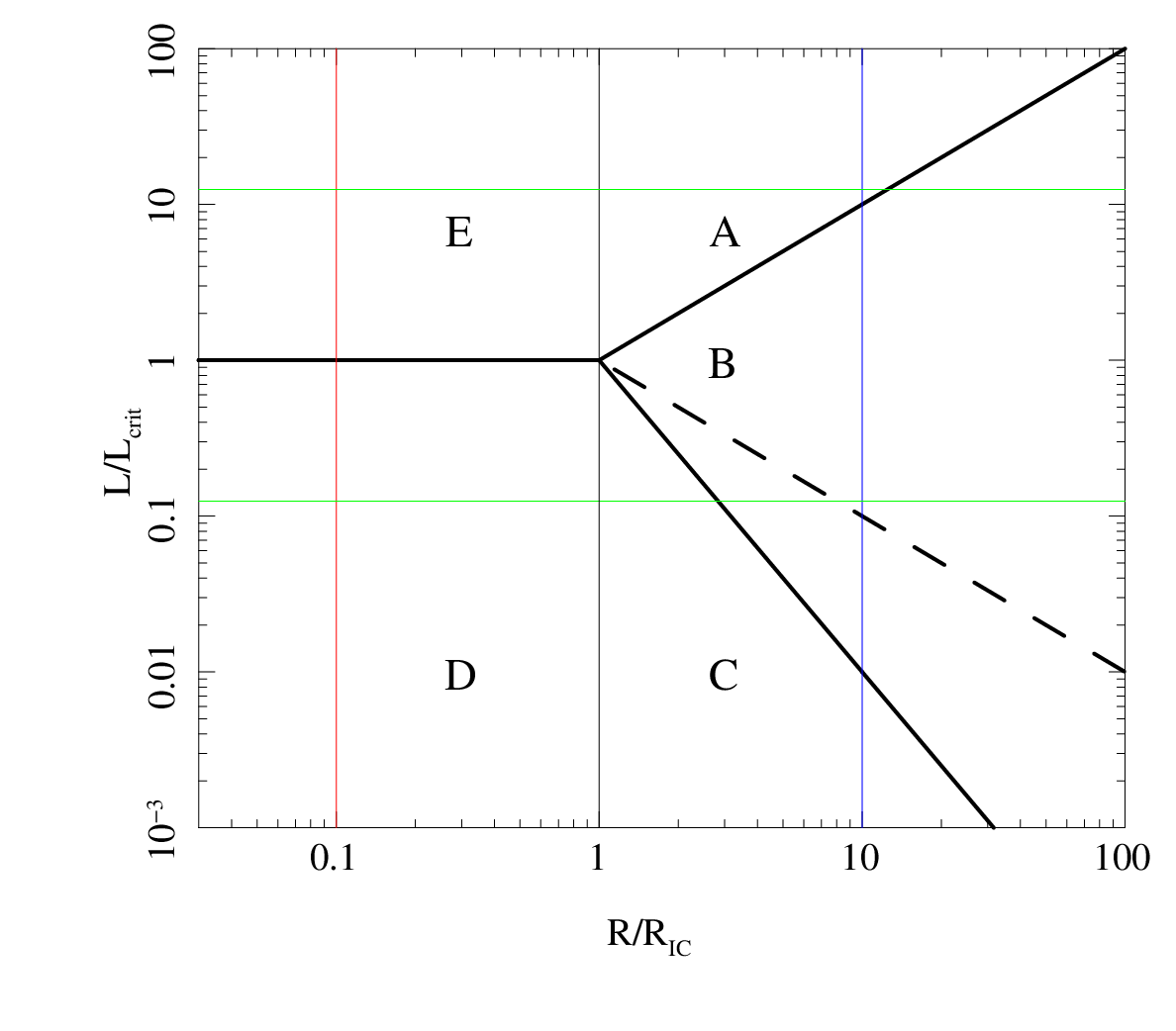} 
\end{tabular}
\caption{The different wind regions of B83, with boundaries shown as
  the thick black lines. For $L>L_{crit}$ the wind is impulsively
  heated to $T_{IC}$ at $R/R_{IC}<L/L_{crit}$ (e.g. upper green
  solid line).  The mean particle velocity is smaller than the escape
  speed for small radii ($R/R_{IC}<1$) so the heated skin forms an
  isothermal atmosphere (region E) with only small wind losses,
  whereas it escapes as an isothermal wind for
  $1<R/R_{IC}<L/L_{crit}$ (region A). For $R/R_{IC}>L/L_{crit}$
  the wind is steadily heated, only reaching temperature $T_{ch}(R)$,
  but this can still escape as a wind (region B) for $L/L_{crit}>1$.
  Instead, for $L/L_{crit}<1$ (e.g. the lower green line), the
  illumination only heats the disc to $T_{ch}$ rather than $T_{IC}$,
  so there is even less wind escape from the inner region with
  $R/R_{IC}<1$ (region D). The material still cannot escape at $R=R_{IC}$ 
as $T_{ch}<T_{IC}$ (region C, gravity inhibited wind), and only becomes unbound for
  $R/R_{IC}\ge (L/L_{crit})^{-1}$ (region B again), though this inner
  boundary depends somewhat on the Compton temperature, with the
  dashed (solid) thick black line being appropriate for low (high)
  Compton temperatures (W96). The red vertical line shows that at
  small radii, $R_{out}=0.1R_{IC}$, the disc will go from region E and
  D as the luminosity increases, whereas the blue line shows that at
  large radii, $R_{out}=10R_{IC}$, the disc material goes from being
  inhibited by gravity (region C), to being steadily heated (region
  B), to impulsively heated (region A) as the luminosity increases.
}
\label{fig:regions}
\end{figure}

\begin{figure*}
\begin{tabular}{cc}
\epsfxsize=0.45\textwidth \epsfbox{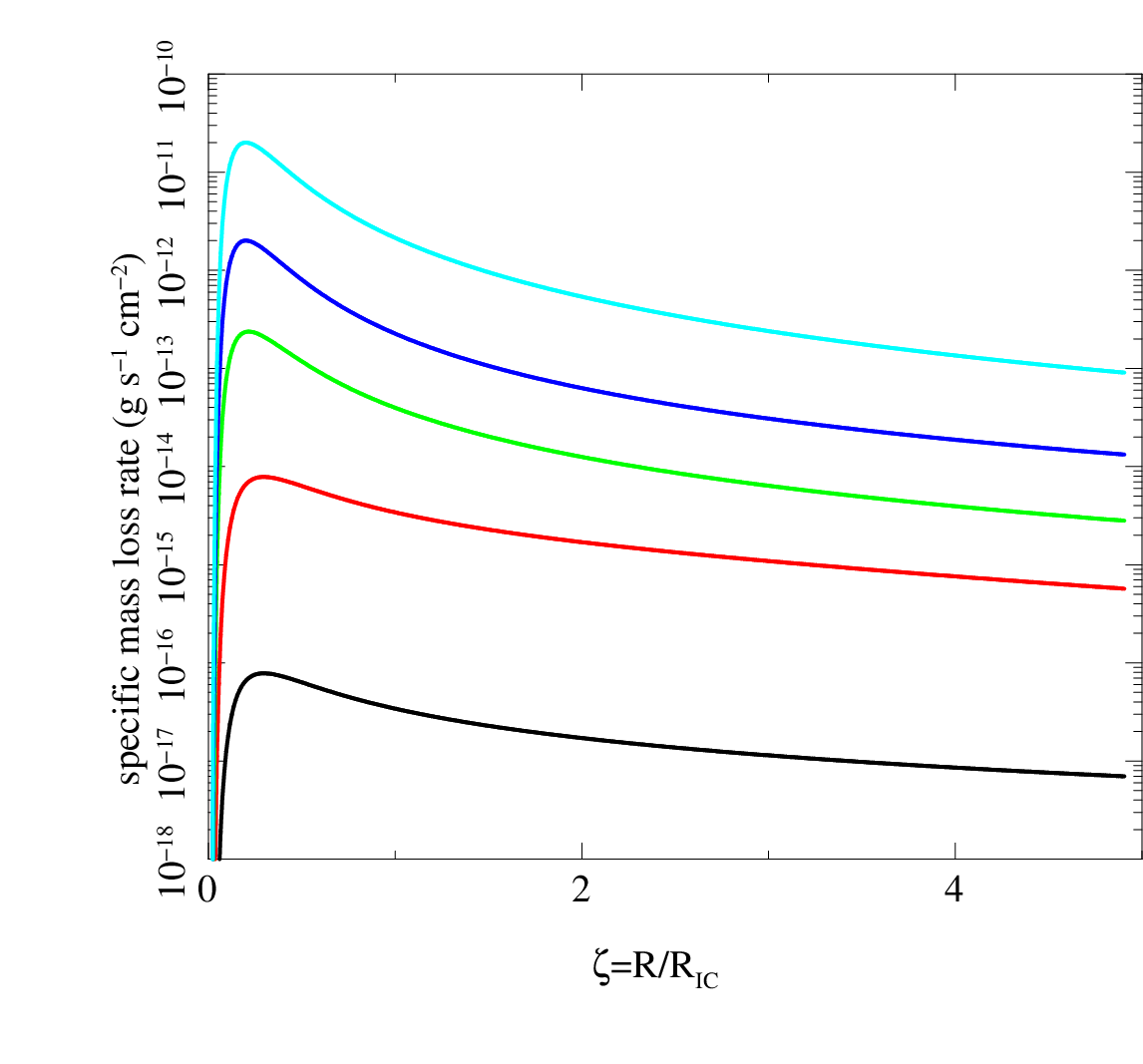} &
\epsfxsize=0.45\textwidth \epsfbox{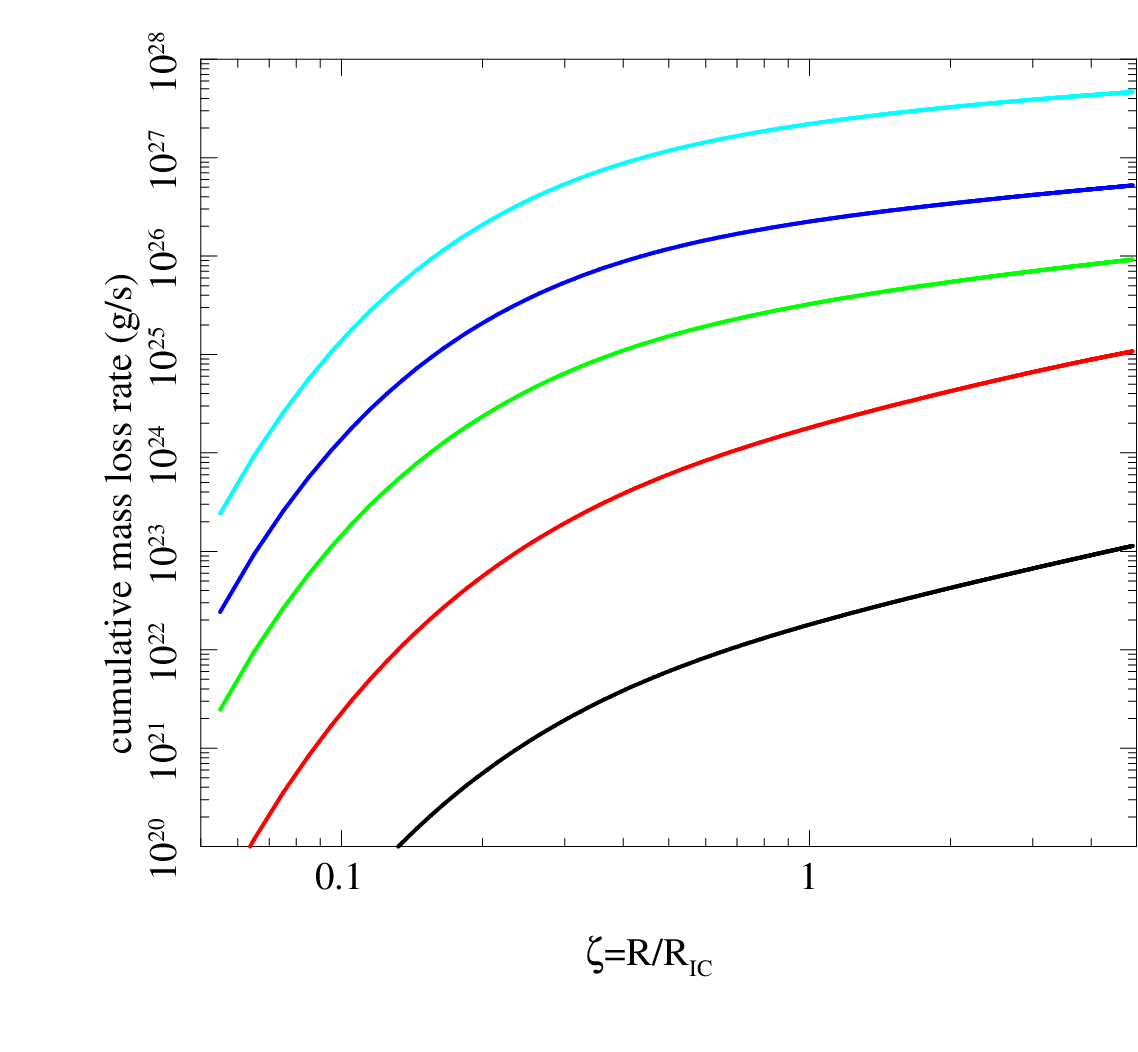}\\
\end{tabular}
\caption{ a) The mass loss rate from a $10^8 M_\odot$ black hole per unit
  area of disc at scaled radius $\zeta=R/R_{IC}$ for $\log
  L/L_{Edd}=-3$ (black) $-2$ (red), $-1$ (green), $0$ (blue) and $1$
  (cyan) as in W96. These correspond to $L/L_{crit}=0.0125-125$ for
  the assumed constant $T_{IC,8}=0.13$. b) The corresponding
  cumulative mass loss rate from the disc at $R<R_{out}=5R_{IC}$. 
}
\label{fig:woods}
\end{figure*}

\begin{figure}
\begin{tabular}{c}
\epsfxsize=0.45\textwidth \epsfbox{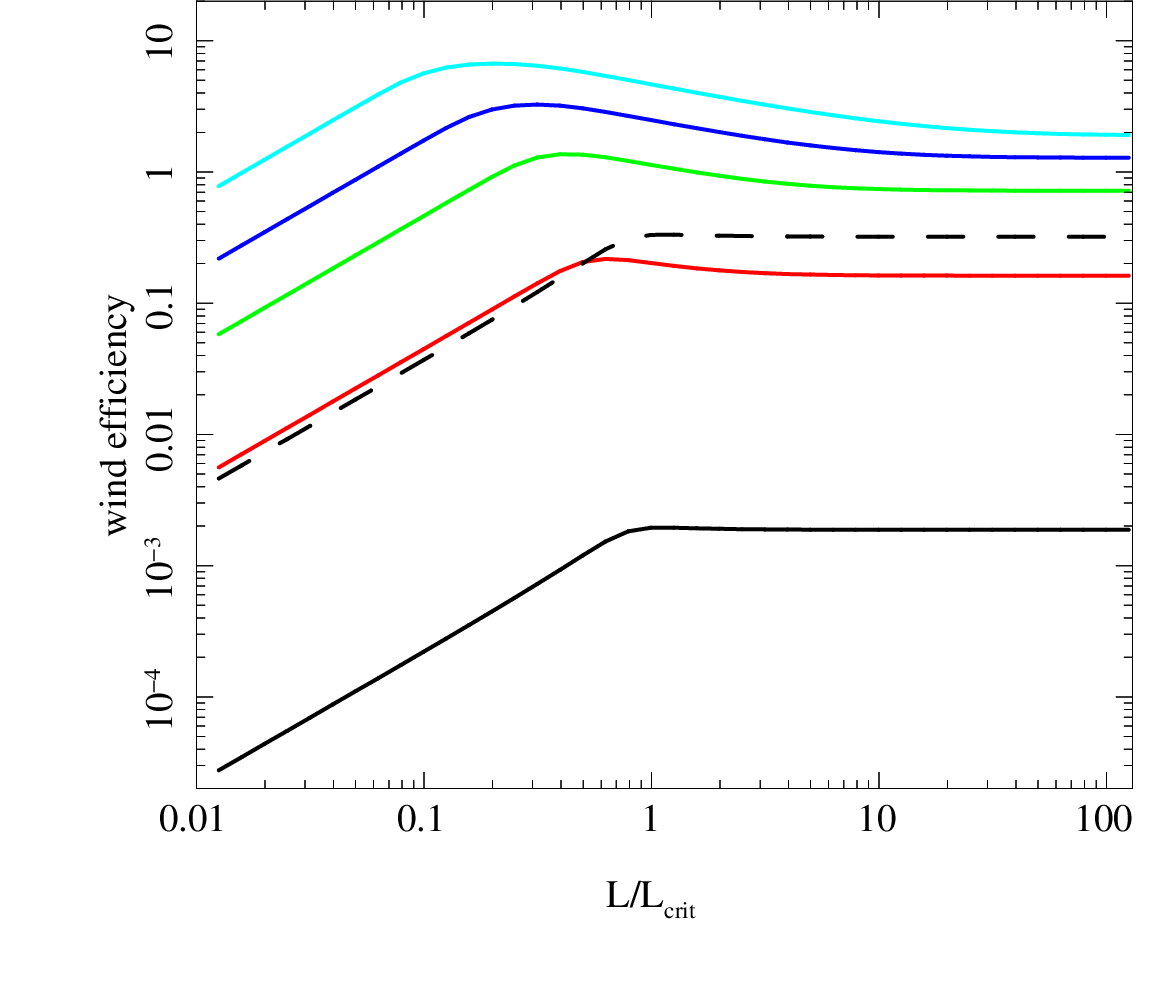} 
\end{tabular}
\caption{ The effect of disc size on the efficiency of wind production
  per unit mass accretion rate.  The range of irradiating luminosity
  is the same as in Fig. 1 i.e. $L/L_{Edd}=10^{-3}\to 10$, which
  corresponds to $L/L_{crit}=1.25\times 10^{-2}\to 1.25\times 10^2$
  for the assumed $T_{IC,8}=0.13$. The disc size is
  $R_{out}=0.1R_{IC}$ (black), $0.3R_{IC}$ (red), $R_{IC}$ (green),
  $3R_{IC}$ (blue) and $10R_{IC}$ (cyan). All disc sizes show that the
  disc is increasingly efficient at producing a wind with increasing
  mass accretion rate for $L/L_{crit}\to 1$, but then the efficiency
  stabilises to a constant for $L/L_{crit}\gg 1$. This constant value
  is larger than the mass input rate for all disc sizes where
  $R_{out}\ge R_{IC}$, so the wind losses should have a major impact
  on the observed luminosity.  B83 suggest that the mass loss rates from the smallest
  disc sizes, with $R_{out}\ll R_{IC}$, are underestimated. The dashed black line shows
  the impact of their suggested correction factor for small discs
  ($R_{out}=0.1R_{IC}$) where there is no confining pressure from an
  outer wind. }
\label{fig:woods_rout}
\end{figure}

The critical luminosity can be written in terms of the Eddington limit
$L_{Edd}=4\pi GM m_I c/\sigma_T$, where $m_I\approx 2\mu$ is the mean ion
mass per electron, giving 
\begin{equation}
\begin{split}
L_{crit}=\frac{1}{8} (m_e/\mu)^{1/2} (m_ec^2/kT_{IC})^{1/2} L_{Edd}\\
\sim 0.03T_{IC,8}^{-1/2} L_{Edd}
\end{split}
\label{eqn:lcrit}
\end{equation}
Thus the entire outer disc out to $R_{out}$ remains impulsively heated
only for $L>L_{out}=(R_{out}/R_{IC})L_{crit}$ (region A in B83, see
Fig.~\ref{fig:regions}). For $L_{crit}<L<L_{out}$ this wind is only
impulsively heated from $R_{in}\sim R_{IC}$ to to a radius
$(L/L_{crit})R_{IC}$. For larger radii the heating is not fast enough
to heat it to $kT_{IC}$, so it only reaches $kT_{ch}$, but gravity is
lower at these larger radii, so it still escapes as a wind as
$kT_{ch}\ge GM\mu/R$ (upper Region B in B83).
The upper horizontal thin (green) line on 
Fig.~\ref{fig:regions} shows an example for
$L/L_{crit}=12.5$, where the wind is impulsively heated for $1<
R/R_{IC}<12.5$ and then steadily heated for radii $R/R_{IC}>12.5$.

Similarly, for $L<L_{crit}$ the wind is only heated to a temperature
of $T_{ch}<T_{IC}$ at $R_{IC}$, so cannot escape (Region C in B83, the
gravity inhibited wind). This escape condition $kT_{ch}>GM\mu/R$,
defines an effective inner launch radius for the wind of $R_{in}/R_{IC}\sim
(L/L_{crit})^{-2}$. This is appropriate at high Compton
temperatures, $T_{IC,8}\sim 1$, where only Compton processes are
important as assumed in B83, but at lower Compton temperatures then
atomic cooling become important, giving $R_{in}/R_{IC}\sim
(L/L_{crit})^{-1}$ (W96, Higginbottom \& Proga 2015).  Either way, this is again the steady
heating region (Region B).  The lower horizontal thin (green) line on
Fig.~\ref{fig:regions} shows an example for $L/L_{crit}=0.125$, where
the wind is launched only for radii $R/R_{IC}>12.5$
(low Compton temperature), or $>2.82$ (high Compton temperature).

The instability is still triggered at $\Xi\sim 40$, giving the density
as $n_{ch} \propto L/(R^2 T_{ch}$), so the specific mass loss rate in region B is
$\dot{m} = n_{ch}\mu c_{ch}$ (no factor of $1/2$ this time, see B83), giving a
total mass loss rate in the wind of
\begin{equation}
\begin{split}
\dot{M}_{wind}=\int_{R_{in}}^{R_{out}} n_{ch}\mu c_{ch} \times 2\times 2\pi R dR\\
=  3\frac{L^{2/3}L^{1/3}_{crit}/c }{40 c_{IC}} 
\Bigl[ \Bigl( \frac{R_{out}}{R_{IC}} \Bigr)^{1/3} - 
       \Bigl( \frac{R_{in}}{R_{IC}} \Bigr)^{1/3} \Bigr]
\end{split}
\label{eqn:mdotb}
\end{equation}
Thus the wind mass loss rate in region B scales only with $L^{2/3}$
rather than with $L$ as in region A.

For $R\ll R_{IC}$ and $L>L_{crit}$ the upper layers of the disc are
heated to $T_{IC}$ but this is less than the escape velocity at this
point. Thus the material forms an isothermal, exponential atmosphere
above the disc with only a very small fraction of the material being
able to escape (region E). At lower luminosities ($L<L_{crit}$) this
atmophere is steadily heated to characteristic temperature
$T_{ch}<T_{IC}$, so there is even less wind (region
D). Fig.~\ref{fig:regions} shows these regions schematically (B83,
W96).

Regions A and B (impulsive Compton heating and steady Compton
heating) are the two main regimes which contribute to the thermal wind,
though other radii/luminosities ranges can power smaller mass loss
rates. W96 give fitting formula for the specific mass loss rate in
all regimes of luminosity and Compton temperature of the radiation in
their equation 4.8. These were derived from smoothly matching to results from
full hydrodynamical simulations of thermal winds.  We integrate these
over all radii, rather than 
using the different equations \ref{eqn:mdota} and \ref{eqn:mdotb}
with upper and lower radial limits. We 
show the results for the specific mass loss rate per unit area,
$\dot{m}$ as a function of radius in Fig.\ref{fig:woods}a, reproducing
the results in Fig 6 of W96 for a $10^8$\Msol black hole with $\log
L/L_{Edd}=-3,-2,-1,0,1$, which is equivalent to $L/L_{crit}=1.25\times
10^{-2}-1.25\times 10^2$ for the assumed $T_{IC,8}=0.13$. Obviously, radiation
pressure should also enhance the wind as the luminosity approaches
and exceeds $L_{Edd}$, which was not considered in W96 or B83. We return to
this point in Section 4. There is a clear peak in the 
specific mass loss rate at  $R\sim 0.2R_{IC}$. 

Fig.\ref{fig:woods}b shows the corresponding cumulative mass loss in
the wind, $\dot{M}$ as a function of $R/R_{IC}$. This shows that the
total mass loss rate in the wind rises quickly at $R\sim 0.2R_{IC}$,
and then increases more slowly with increasing $R$. This can be
understood from the previous plot of specific mass loss rate, as this
declines as $R^{-2}$ in the wind regions A and B. Hence the
increasing area at larger distances means that the total mass loss
rate from the disc increases with increasing size scale of the disc,
as $\dot{M}\propto \log R_{out}/R_{IC}$ for $R>3R_{IC}$. For
$L<L_{crit}$ there is a steeper dependence as a larger fraction of the 
wind is launched from further out in the disc.

Fig.\ref{fig:woods_rout} shows the efficiency of wind production,
defined as the ratio of mass loss rate in the wind to the mass
accretion rate required to produce the wind (which is $\propto L$). We
show this efficiency as a function of $L/L_{crit}$ for different disc
sizes, with $R_{out}=0.1R_{IC}$ (black), $0.3R_{IC}$ (red) $1R_{IC}$
(green), $3R_{IC}$ (blue), and $10R_{IC}$ (cyan).  For large discs,
there is a clear change in behaviour from constant efficiency (wind
mass loss rate scaling with $L$, region A) at high $L/L_{crit}$ to
increasing efficiency at lower $L/L_{crit}$ (wind mass loss rate
scaling with $L^{2/3}$ in region B) for $L_{min}< L<L_{crit}$ before
the steep decline in efficiency for $L<L_{min}$ (region C, see (blue)
vertical line in Fig.~\ref{fig:regions} for
$R_{out}=10R_{IC}$). Thus the minimum luminosity to produce an
efficient wind is $L_{min}= L_{crit} (R_{IC}/R_{out})$ (low Compton temperature)
rather than $L>L_{crit}$. Most of the wind is launched from the outer disc
(see Fig.\ref{fig:woods}b) so the minimum requirement for a wind is that the
outer radii of the disc are in region B, rather than that the wind
extends down to $R_{IC}$. 

Thus thermal winds are predicted to be strongest in the systems with
largest discs.  This reflects the observed distribution. Winds are
preferentially seen from systems with the longest orbital periods
i.e. the systems with the largest discs (Diaz-Trigo \& Boirin
2016). It is highly unlikely that a magnetic wind would depend on the
outer disc size, so this alone favours a thermal wind origin to
explain the observed data.

Instead, for a small disc with $R_{out}\ll
R_{IC}$ the wind changes from constant but low efficiency (exponential
scale height atmosphere with isothermal temperature, region E) to
being strongly suppressed by inefficient heating in region D (red
vertical line in Fig.~\ref{fig:regions} for a disc with
$R_{out}=0.1R_{IC}$).

\begin{figure*}
\begin{tabular}{cc}
\epsfxsize=0.45\textwidth \epsfbox{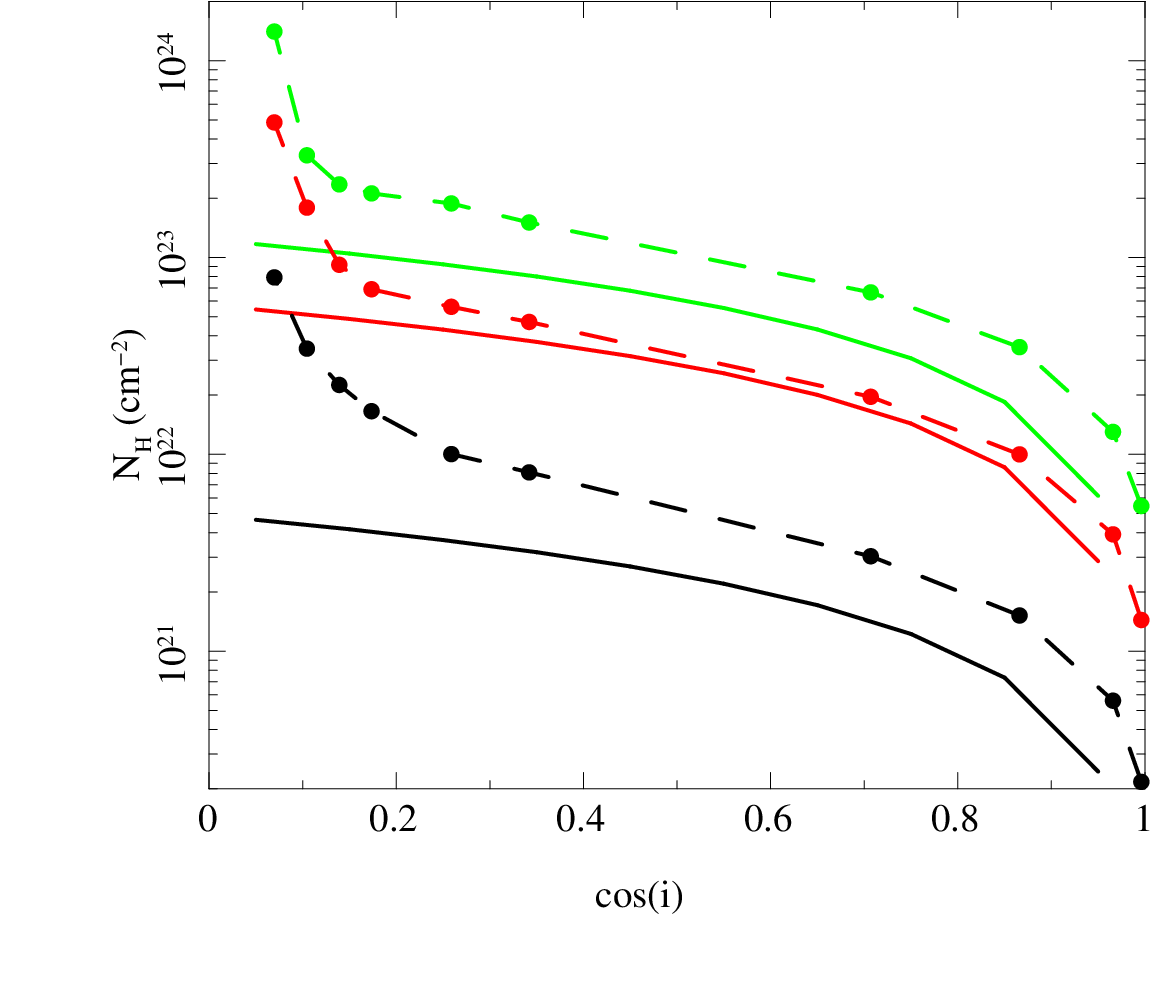} &
\epsfxsize=0.45\textwidth \epsfbox{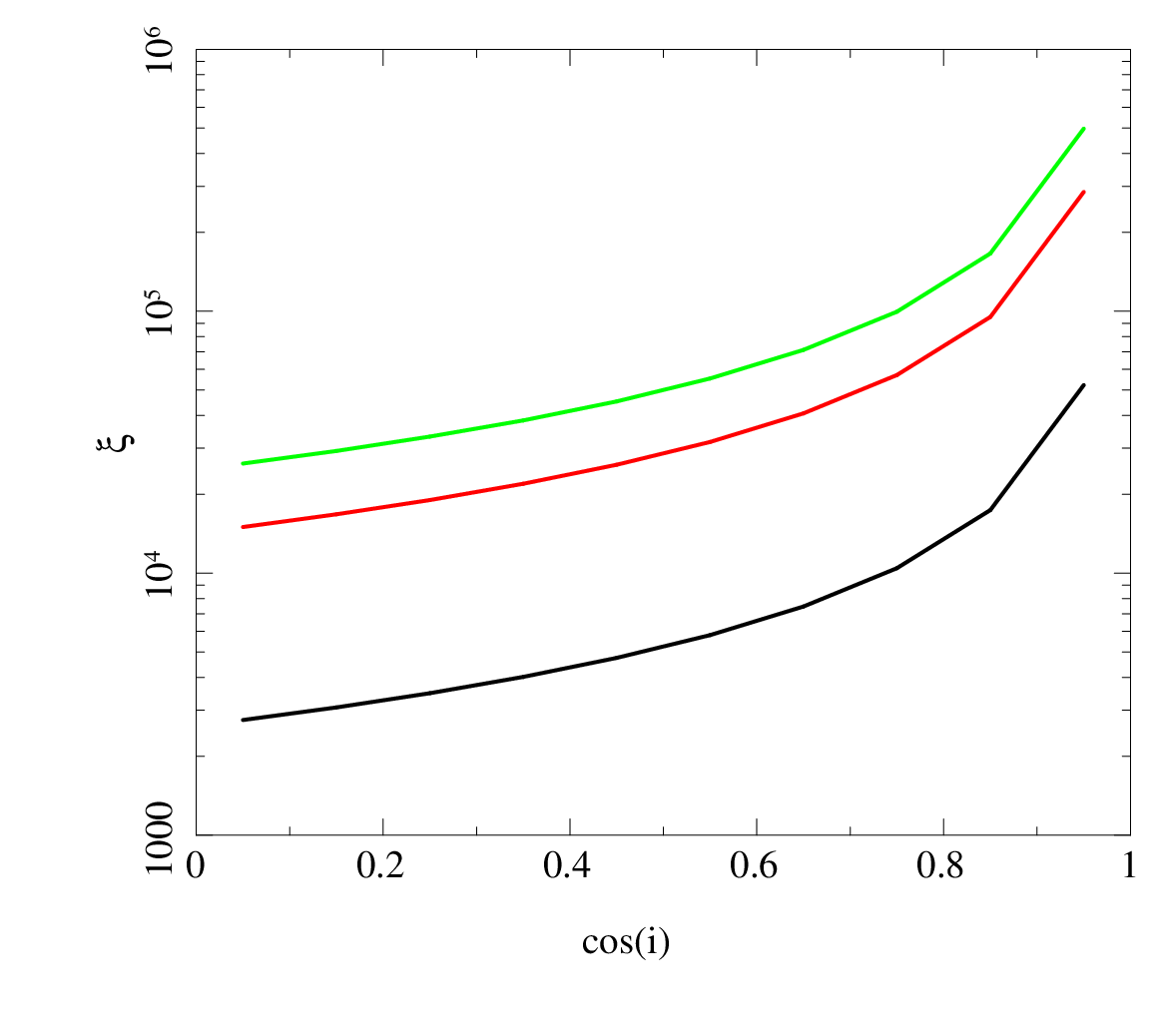}\\
\end{tabular}
\caption{a) The dashed line/open circles show the column density in
  the wind (excluding the static corona) as a function of cosine of
  the inclination angle from the hydrodynamic simulations of W96 for
  $T_{IC,8}=0.13$ for pairs of $L/L_{Edd}, R_{out}$ of $0.01, 12$
  (black) $0.08, 5$, (red) and $0.3, 5$ (green). The solid lines show
  the columns resulting for each of these from the analytic
  approximations assuming that the density is $\propto (1-\cos(i) )$ and
  that the typical velocity of the wind is the average (mass loss
  weighted) sound speed. This is $\sim 140, 350$ and $420$ km/s for
  each of the simulations, respectively.  The very simple analytic
  approximations agree with the results of the hydrodynamic
  simulations to within a factor $\sim 1.5-2$ over a broad range of
  angles.  b) The ionisation parameter of the wind.  This decreases at
  higher inclination (lower $\cos(i)$) as the density in the wind is
  higher. }
\label{fig:nh}
\end{figure*}

The wind efficiency is larger than unity for all discs with
$R_{out}>R_{IC}$ and $L>L_{min}$, showing how powerful these winds
can be. This is a feature of thermal winds which was stressed in
the early papers, that they can remove more mass than is required to
keep the source bright enough to power the wind (B83).  Shields et al
(1986) showed that these strong wind losses could cause oscillations
in an otherwise steady state disc. However, most black hole binaries,
especially those with large discs, are unstable to the hydrogen
ionisation disc instability (see e.g. Lasota 2001). Thus the strongest
wind losses are not likely to be in stable disc systems. Instead, the
wind mass loss should impact on the length of the outburst in
transient black hole binary sources, as noted by Dubus et al (2001).

B83 comment that the mass loss rates are probably
underestimated for the smallest disc sizes ($R_{out}<R_{IC}$) due to
the changing streamline pattern that results from not having an outer
disc wind to collimate the flow from the inner regions.  They suggest
that this leads to an increase in the wind mass loss rate by a factor
$1/\zeta^2$.  We show the effect of this for the lowest
$R_{out}=0.1R_{IC}$ (i.e. $\zeta\le 0.1$) as the dashed black line in
Fig.\ref{fig:woods_rout}. Plainly, this correction has a very large
impact on the mass loss rate predictions from small discs, but it is
highly uncertain as there are no hydrodynamical simulations in this
regime. Hence we do not consider this further here. 

%==============================================
\subsection{Predicting the absorption structures}
%==============================================

Fig.\ref{fig:woods_rout} shows that the thermal wind drops
dramatically when the luminosity drops below the minimum required to
heat the material on the outer disc to the escape velocity.  This
minimum luminosity is a few percent of the Eddington limit for
$T_{IC,8}=1.0$ for a disc with $R_{out}\sim R_{IC}$. Co-incidentally,
this is also the point at which the accretion flow makes its dramatic
transition from a soft (disc dominated) to a hard (Compton dominated)
spectrum. Thus it is possible that the loss of the wind in the hard
state is simply a consequence of the lower luminosity in this state
leading to a heating rate which is not sufficient for the wind to
reach its Compton temperature. However, the large spectral change also
means that there is a large change in Compton temperature, with the
hard spectrum having larger $T_{IC}$, so $L_{crit}$ and $R_{IC}$ both
go down. This illustrates the difficulty with understanding how the
wind responds when the equations are written in terms of variable
parameters $R_{IC}$ and $L_{crit}$ which depend implicitly on $T_{IC}$
rather than absolute $L$ and $R$.

The key observables are the column density in the wind, and its
ionisation state. The velocity is also another potential observable
but this requires excellent spectral resolution to accurately
determine the low velocities expected from an outflow from the outer
disc. These are currently only accessible using the Chandra HETGS
(especially the 3rd order grating data: Miller et al 2015), but
should become much better determined in the future with calorimeter
resolution. To derive these observables from the models really
requires the full information on the wind density and velocity as a
function of height at all radii. This is beyond the scope of the
analytic equations discussed above. However, these are included in the
hydrodynamic simulations of W96, and they show column as a function of
angle for three of their simulations for $L/L_{Edd}=0.3,0.08$ and
$0.01$ for fixed illuminating spectrum with $T_{IC,8}=0.13$ and disc
$R_{out}=5R_{IC}$ (except for $L/L_{Edd}=0.01$, where
$R_{out}=12R_{IC}$: Woods, private communication).  W96 give the total
column (their tables 2-4), but this includes the column density from the static corona
region as well as the wind, whereas our mass loss rates are (by
definition) for the wind alone. Hence we use their estimate for the
fraction of the column which arises in the static corona (below the
softening radius) to derive the column from the wind material in order
to compare with the analytic equations. The resulting wind column
depends on inclination angle, $i$, which can be approximately
described as $N_H(i)\propto (1-\cos(i) )$ for all angles which do not
intersect the disc photosphere itself.

We make the highly simplified assumptions that the velocity is
constant, given by the mass loss weighted average sound speed.  This
gives $v_{out}$ of $\sim 140, 350$ and $420$~km/s for
$L/L_{Edd}=0.01,0.08$ and $0.3$, respectively.  We also define an
inner radius for effective launching of the wind
i.e. $R_{in}=0.2R_{IC}$ for $L\ge L_{crit}$ (Region A) and $R_{in}
=0.2R_{IC}/(L/L_{crit})$ for $L< L_{crit}$ (using the low temperature
prescription for Region B according to W96 rather than B83, dashed
line in Fig.~\ref{fig:regions}). We assume that the wind streamlines
are radial, that $n(R,\cos(i))=n_0(R)(1-\cos(i))$, and use mass
conservation to fix the edge-on density $n_0(R)=\dot{M}/(4\pi R^2
v_{out} m_i)$. This gives a column of
\begin{equation}
N_H(i)=\int_{R_{in}}^{R_{out}} 
n_0(R)(1-\cos(i)) dR = \frac{\dot{M}(1-\cos(i))} {4\pi R_{in}v_{out} m_i}
\label{eqn:nh}
\end{equation}
We show this prediction (solid lines) overlaid on the hydrodynamic
wind results of W96 (points joined by dashed lines) in
Fig.\ref{fig:nh}, using their assumed parameters ($T_{IC,8}=0.13$,
$M=10^8M_\odot$ and $R_{out}=5R_{IC}=7.5\times 10^{18}$~cm, or $12R_{IC}$ for $L/L_{Edd}=0.01$). Our
results are within a factor of 2 of the hydrodynamic simulations, 
which is a remarkable match given the simplistic assumptions. 

The standard photon ionisation parameter $\xi=L_{ion}/(nR^2)$, where
$L_{ion}$ is the ionising luminosity $\approx L$ for black hole binaries as almost all
the luminosity is emitted above 13.6~eV. The ionisation parameter is 
constant along the radial streamlines, given by 
\begin{equation}
\xi(i)=\frac{4\pi m_i v_{out} L} {\dot{M} (1-\cos(i))}
\label{eqn:xi}
\end{equation}
Hence the ionisation state is lower close to the orbital plane ($\cos(i)=0$),
where the density of the wind is higher, and lower for lower luminosity. 
These are shown in Fig.~\ref{fig:nh}b. 

Thus the thermal wind models can predict not only the overall mass
loss rate from the disc, but also the zeroth order observables of column density,
and mean ionisation state and velocity along any sightline for any
disc size, luminosity and for any radiation spectrum.
However, current data  (e.g. Miller et al 2015)
already give better resolution on the structure of these winds than our simplistic, single zone 
model, and we stress that full hydrodynamic simulations of thermal winds are required for a detailed
comparison.

%==============================================
\section{Evolution of the wind with $L/L_{Edd}$}
%==============================================

Most black hole binaries are transient, showing outbursts in which the
mass accretion rate onto the central object changes dramatically due
to the Hydrogen ionisation disc instability (see e.g. the review by
Lasota 2010). There is an abrupt transition in spectral state on the
fast rise from a hard spectrum which can be roughly described by a
power law with photon index $\sim 1.6-2$, to a much softer spectrum
which is dominated by a multi-colour disc component.  This
hard to soft transition is not at a well defined luminosity, most
probably because the mass accretion rate is changing too rapidly for
the disc to be in steady state (Smith et al 2002; Gladstone, Done \& Gierlinski
2007). Instead, the slow decline is more stable, with the spectrum
changing back from a disc to power law spectrum at $L\sim 0.02L_{Edd}$
(Maccarone 2003). During the disc dominated, most luminous phase, the
characteristic disc temperature is $kT_{max}\propto
(L/L_{Edd})^{1/4}$. The outer disc sees this at high inclination, so
the Doppler blueshift increases the temperature seen by the disc. We
model a $10M_\odot$ black hole with spin $a_*=0.5$ 
using the {\sc kerrbb} code in {\sc xspec} which has full general
relativistic emissivity and ray tracing. Assuming a mass accretion
rate of $3.5\times 10^{17}$~g~s$^{-1}$ ($L/L_{Edd}\sim 0.02$), with a
colour temperature correction of $f_{col}=1.7$, the outer disc sees a
spectrum similar to a multicolour disc blackbody with maximum
temperature of $0.6$~keV, which 
corresponds to $kT_{IC}=0.31$~keV which is $0.036\times
10^8$~K. Hence this predicts that in the disc dominated spectra,
\begin{equation}
T_{IC,8}=0.036 [L/(0.02L_{Edd})]^{1/4}
\end{equation}
 
The transition to the hard state is complex, with the disc temperature
decreasing rapidly, as expected if the thin disc starts to recede from
the innermost stable circular orbit (e.g. Gierlinski, Done \& Page
2008). Here we assume that the spectrum abruptly changes to a power
law of photon index $\Gamma=2$ at $L=0.02L_{Edd}$, flattening to
$\Gamma=1.6$ at $L/L_{Edd}=2\times 10^{-3}$. Interpolating logarithmically
$\Gamma=1.6+0.4\times \log_{10}[L/(0.002L_{Edd})]$. The 
Compton temperature for a hard power law depends on the high
energy cutoff, but this dependence saturates above 
100~keV due to the 
rollover in the Klein-Nishima cross-section compared to the constant
cross-section assumed in Thomson scattering. Hence we fix the upper limit
of the flux integral at 100~keV, and assume a lower limit of 0.1~keV. 
This gives an inverse Compton temperature of
3.6~keV ($0.42\times 10^8$~K), increasing to 7.6~keV ($0.88\times
10^8$~K) for the hardest spectra/lowest luminosities considered
here, so that
\begin{equation}
T_{IC,8}=0.88 - 0.46\times \log_{10}[L/0.002L_{Edd}]
\end{equation}

We use this correlated change in $kT_{IC}$ with $L/L_{Edd}$ to explore
the predicted wind behaviour over the range $10^{-3}L_{Edd} <L<
L_{Edd}$, with the total mass loss rate in the wind $\dot{M}$
calculated from the fitting formulae of W96 and the column/ionisation
state observables approximated as in Section 2 above. We assume a
generic black hole of mass $M=10M_\odot$ and set the disc outer radius
to $3.7\times 10^{12}$~cm ($2.5\times 10^6 R_g$, which corresponds to $5R_{IC}$ for
$T_{IC,8}=0.13$ as used in most of the simulations in W96).  We
note that most BHB have much smaller discs, but the most dramatic
winds are indeed seen in systems which are known to be in long period
orbits (GRS1915+105, GRO J1655-40), or where the orbital periods are
unknown but consistent with being long (H1743-322, 4U1630-522)
e.g. Diaz Trigo \& Boirin (2016).

\begin{figure}
\begin{tabular}{c}
\epsfxsize=0.45\textwidth \epsfbox{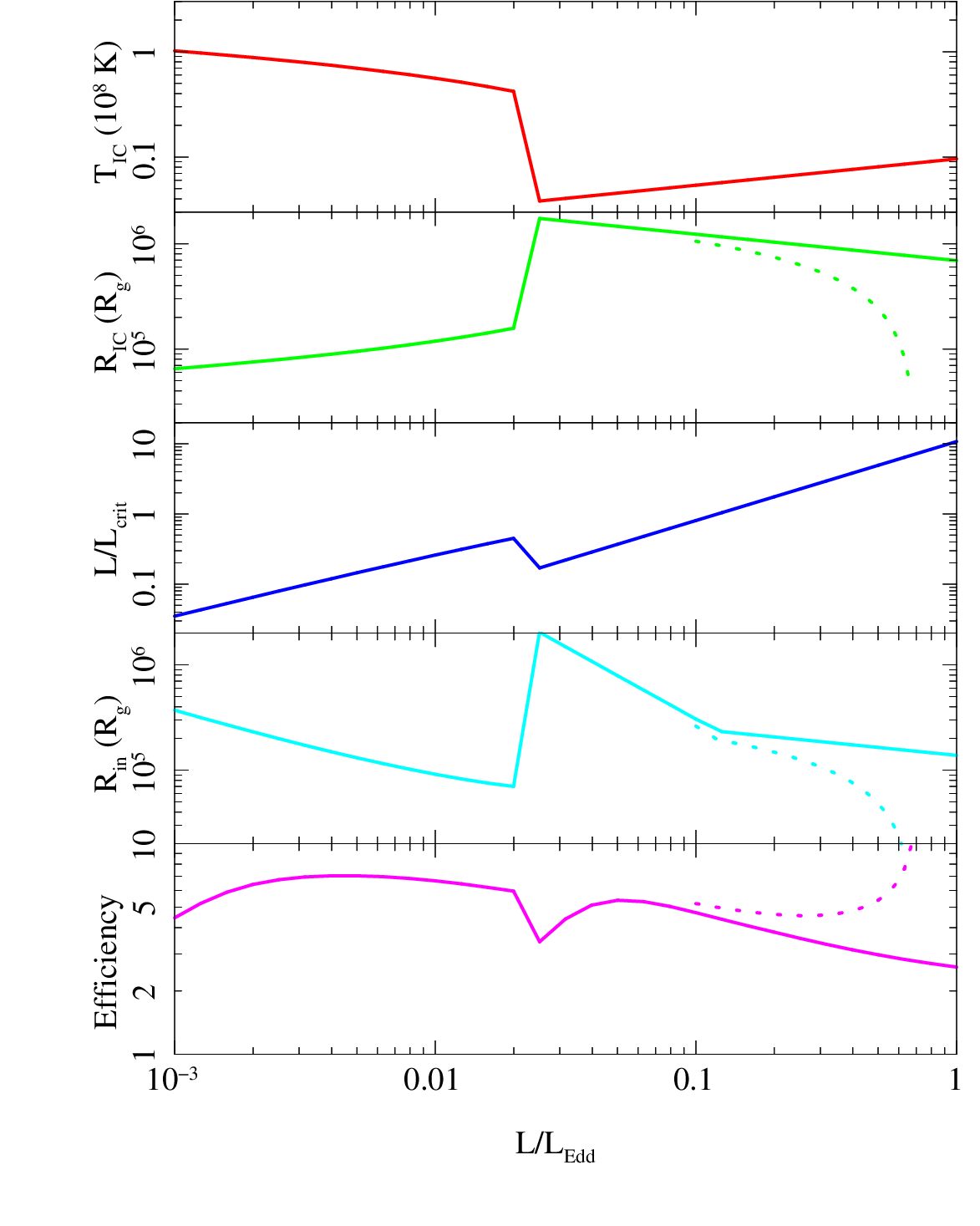} \\
\end{tabular}
\caption{
The upper panel shows the assumed change in $T_{IC}$ with $L/L_{Edd}$
(red).  The power law spectral index softens with increasing
luminosity in the low/hard state, so the inverse Compton temperature
drops. The abrupt drop at $L/L_{Edd}=0.02$ marks the transition to the
disk dominated state, where the Compton temperature increases with
luminosity. The second panel shows the effect of this on $R_{IC}$
(green). A higher Compton temperature means that the wind can escape
from smaller radii. The third panel shows how $L/L_{crit}$ changes
(blue). An increase in Compton temperature means that the radiation
heats the gas faster, so it can drive the temperature up to $T_{IC}$
at $R_{IC}$ at a lower luminosity. The assumed $L/L_{Edd}-T_{IC}$
behaviour means that the luminosity is only above the critical
luminosity for $L>0.1L_{Edd}$. Thus the wind is lauched from
$R_{in}=0.2R_{IC}$ only for $L>0.1L_{Edd}$. Below this, the wind
instead is launched from $R_{in}=0.2R_{IC}/(L/L_{crit})$, so the wind
launch radius (cyan) decreases with increasing $L/L_{Edd}$ in the
low/hard state, as the increase in $L/L_{crit}$ more than offsets the
increase in $R_{IC}$. The lower panel (magenta) shows the wind efficiency 
(mass outflow rate in terms of the mass accretion rate). This is 
fairly constant at $\sim 5\times$ the mass accretion
rate required to produce the luminosity, except for more complex
behaviour around the transition.  The dotted lines on all panels show
the effect of including a simple radiation pressure term to reduce the effective
gravity. The wind can be launched from progressively smaller radii, and the mass
loss rates increase.
}
\label{fig:bhb_over}
\end{figure}

The upper panel of Fig. \ref{fig:bhb_over} shows the Compton
temperature as a function of $L/L_{Edd}$, illustrating the dramatic
change in behaviour at the hard/soft spectral transition at $L\sim
0.02L_{Edd}$.  This has a similarly dramatic impact on the radius at
which Compton temperature allows material to escape, $R_{IC}$, shown
as the solid line in the second panel of
Fig. \ref{fig:bhb_over}. However, the effective launch radius,
$R_{in}$, depends on luminosity relative to the critical luminosity
which is required to launch the wind efficiently, which itself depends
on $T_{IC}$.  The third panel in Fig. \ref{fig:bhb_over} shows
$L/L_{crit}$. This only reaches unity for $L>0.1L_{Edd}$, so only
above this luminosity is $R_{in}=0.2R_{IC}$ (fourth panel,
Fig. \ref{fig:bhb_over}).  Below this, $R_{in}=0.2
R_{IC}/(L/L_{crit})$ so it decreases with increasing luminosity in the
low/hard state, as well as showing more complex behaviour at the
transition due to the jump in Compton temperature. However, all of these
are within the disc outer radius, so the outer parts of the disc are
in the strong wind region B even for $L/L_{crit}<1$. 
The final panel shows the wind efficiency i.e. the ratio of mass loss rate in
the wind with the mass accretion rate required to produce the
assumed luminosity.  This is fairly constant as the outer disc 
is always in one of the strong wind regions, and high at $\sim 5\times$
the input mass accretion rate. This is a 
little larger than in the simulations in W96 and Fig.~\ref{fig:woods_rout}
as we are assuming a black hole with spin $0.5$ rather than spin 0, so
the same luminosity is produced with
a lower mass accretion rate. 

\section{Radiation pressure correction}

Neither B83 nor W96 include radiation pressure on electrons in their
hydrodynamic simulations. However, this must become important as $L\to
L_{Edd}$. By definition, static material above the disc will be driven
out as a wind at $L>L_{Edd}$, as the radiation pressure reduces the
effective gravity to $GM/R^2(1-L/L_{Edd})$. However, the wind material
is not static as it has the Keplarian rotation velocity from where it
was launched as well as thermal motion driven by the pressure
gradients. This will mean that it becomes unbound at all radii at
luminosities somewhat below $L_{Edd}$. A lower limit to this
completely unbound luminosity is $L_{Edd}/2$, which could be reached
if all the Keplarian azimuthal rotational velocity
($v_\phi=v_{esc}/\sqrt{2}$) were converted to radially outward velocity
(Ueda et al 2004). Conserving angular momentum as well as energy 
pushes this up to $L_{Edd}/\sqrt{2}$. This is very close to the results of a full
calculation in Proga \& Kallman (2002). In the case where the disc
luminosity at the wind launch radius is negligible, they show
(equations 20-22) that the effective gravity goes to zero at $H\sim R$
for $L\sim L_{Edd}/(1+\pi/8)\approx L_{Edd}/\sqrt{2}$. This gives a
simple correction to the Compton radius of
\begin{equation}
\overline{R}_{IC}\approx R_{IC} \Bigl( 1-\frac{L}{0.71L_{Edd}}\Bigr)
\end{equation}

We first assume that this correction to the inverse Compton radius,
$\overline{R}_{IC}$, is the only change in wind properties, and rerun
the black hole binary simulation.  The new results are shown as the
dotted lines on Fig.~\ref{fig:bhb_over}. The lower effective gravity
as the luminosity increases towards Eddington means that $R_{IC}$ and
hence $R_{in}$ both decrease dramatically, formally going to zero at
$0.71L_{Edd}$ (dashed green and cyan lines in the second and fourth
panels).  The wind can then be launched from everywhere on the disc,
so the mass loss rates also increase. We note that this is likely to
be an underestimate of the increase in mass loss rate as we still
assume that the wind velocity is given by the sound speed, but it 
should be higher due to the contribution of radiaton pressure to the
acceleration.

\begin{figure}
\begin{tabular}{c}
\epsfxsize=0.45\textwidth \epsfbox{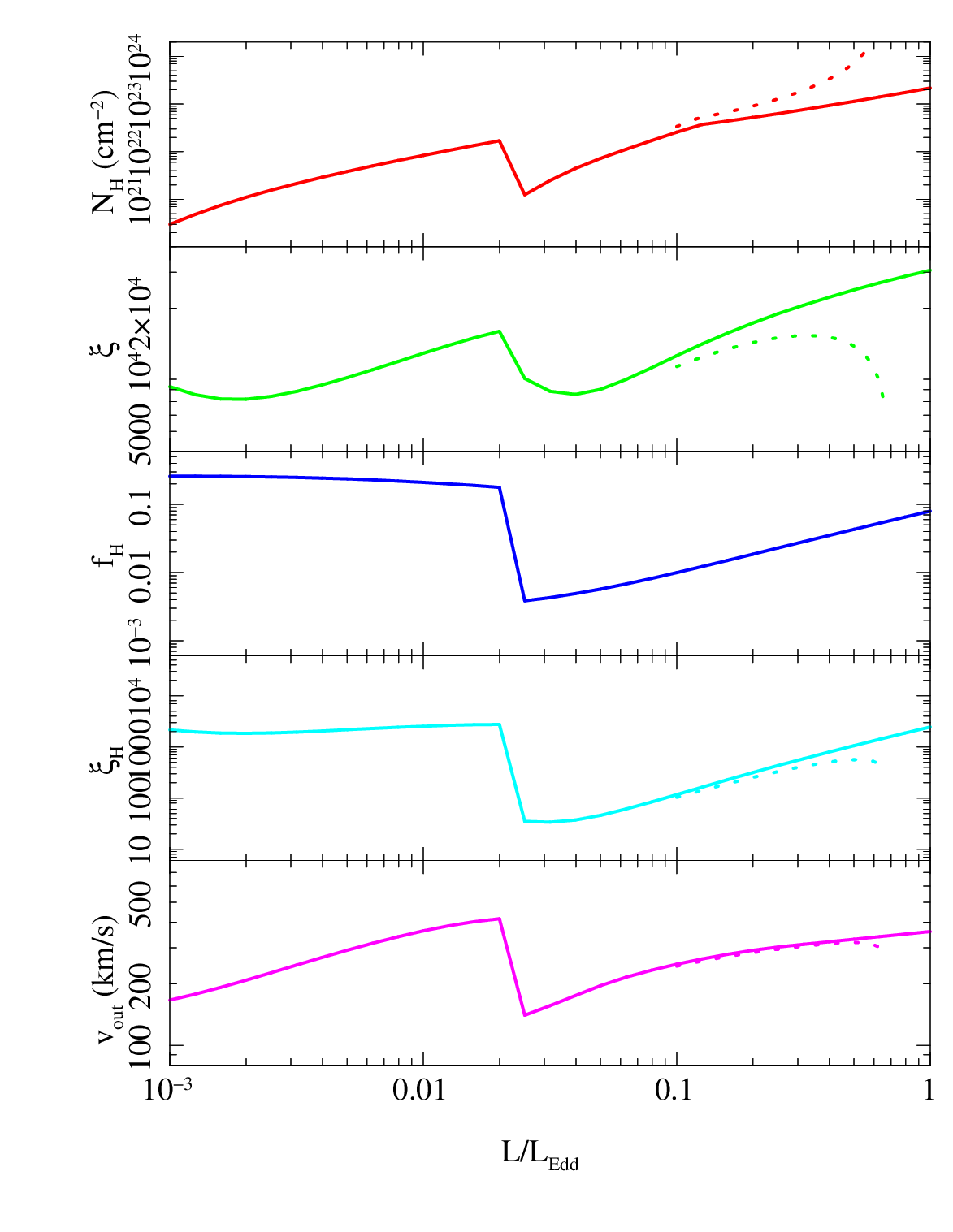} \\
\end{tabular}
\caption{The upper panel (red) shows the column density at $\cos(i)=0.25$
  i.e. an inclination angle of $\sim 75^\circ$. This is roughly
  proportional to the mass accretion rate, but with a dip at the
  spectral transition due to the lower Compton temperature of the
  dimmest high/soft states.  The second panel (green) shows the
  ionisation state calculated from the total bolometric flux,
  $\xi=L_{bol}/nR^2$. This is fairly constant, as the increase in
  luminosity is mostly balanced by an increase in density of the wind.
  The third panel shows the fraction of bolometric flux which is
  emitted in the 8.8-30~keV high energy bandpass (blue). These are the
  photons which are most effective in photoionising He- and H-like
  iron, and this shows a dramatic dip around the spectral transition,
  even including a power law to higher energies which carries 5\% of
  the total disc luminosity.  The fourth panel shows the high energy
  ionisation parameter, $\xi_H=f_H L_{bol}/nR^2$ (cyan). This is almost
  completely constant in the low/hard state, but dips dramatically in
  the high/soft state due to the much softer spectra, and only
  recovers to the same value as seen in the low/hard state at the
  highest luminosities. The bottom panel shows the outflow
  velocity. This mirrors the behaviour of the launch radius of the
  wind shown in Fig.\ref{fig:bhb_over}d. The dashed lines show the
  effect of including a simple radiation pressure correction. The wind
  becomes optically thick ($N_H>2\times 10^{24}$~cm$^{-2}$, and has
  much lower ionisation parameter around $L\sim 0.7L_{Edd}$, as
  required to explain the dramatic wind in GRO J1655-40 (Miller et al
  2006, Shidatsu et al 2016).}
\label{fig:bhb_obs}
\end{figure}

\section{Predicted absorption features}

\subsection{Behaviour at the transition}

We calculate the observable features of the simple model for the BHB
spectral evolution as a function of $L/L_{Edd}$ introduced in the
previous sections.  Fig.\ref{fig:bhb_obs} shows the observables of
$N_H$ (upper panel, red) and $\xi$ (middle panel, green), extracted
for an inclination of $75^\circ$ ($\cos(i)=0.25$).  The solid lines show the standard
thermal wind model results. The column density is
roughly proportional to mass accretion rate, but with a drop at the
spectral transition due to the large change in spectral shape changing
the Compton radius and critical luminosity. This predicts that there
should be an abrupt increase in column by around a factor 10 as
the source declines and makes the transition to the low/hard state. 
This is exactly opposite to the claimed behaviour of the wind shutting off 
in the low/hard state.

However, the visibility of the wind is also controlled by its
ionisation state.  The ionisation parameter calculated from the full
luminosity, $\xi=L/(nR^2)$, is almost constant, changing only by a
factor 5 as the luminosity varies by a factor of a thousand.  This is
because the ionisation is roughly proportional to 
ratio of luminosity and wind mass loss rate
(equation~\ref{eqn:xi}) so these cancel in the regime where the 
wind efficiency is roughly constant. However, the
photo-ionisation of iron depends on the high energy 8.8-30~keV flux,
which changes dramatically at the transition. This high energy flux
would be very small for the lowest luminosity high/soft states as
these have low temperature discs with very few photons emitted above
8.8~keV. However, such pure disc spectra are rare. Most high/soft
states have a small, soft power law tail giving some higher energy
flux.  Hence we also include an additional power law in the high/soft
state, with index fixed at $\Gamma=2.2$, which carries 5\% of the
total power. This has little impact on the Compton temperature, but
will determine the photo-ionisation of iron ions. The third panel in
Fig.\ref{fig:bhb_obs} (blue) shows the fraction of the bolometric flux
which is emitted in the high energy 8.8-30~keV bandpass, $f_H$, while
the fourth panel (cyan) shows the corresponding high energy
photoionisation parameter, $\xi_H=f_H L/(nR^2)= f_H\xi$. This drops by a
factor of more than 50 at the transition, so the photo-ionisation of
the wind changes dramatically. 

However, the baseline ion population of iron is set by collisional
ionisation rather than photo-ionisation - the wind is heated to a
temperature $T_{ch}$, so the ion populations cannot drop below those
which characterise material at this temperature.  We will explore this
in detail in a subsequent paper, but here we just note that neither
the Compton temperature nor the high energy photo-ionisation parameter
are high enough to completely strip iron in the wind in the low/hard
state. Thus while there is a large change in ion populations in the
wind across the transition, it is not likely that this is the origin
of the lack of iron absorption features in the wind in the low/hard
state (Neilsen \& Lee 2009; Miller et al 2012).

\subsubsection{Comparison to observations across the transition}

We analyse the constraints from the data across the transition
in more detail. The
observations of Neilsen \& Lee (2009) are of GRS1915+105, a source
which is close to its Eddington limit, even in its hardest spectral
states, which is a hard intermediate state rather than a true low/hard
state (Done, Wardzinski \& Gierlinski 2004). Hence this requires more
detailed modelling which will be the subject of a later paper
(Shidatsu et al, in preparation). However, the observations of
H1743-322 of Miller et al (2012) are in standard low/hard and
high/soft states which are directly comparable to those simulated in
the previous section. Hence we can use our models to directly compare
to the observed thermal wind features across the spectral transition.

\begin{figure}
\begin{tabular}{c}
\epsfxsize=0.45\textwidth \epsfbox{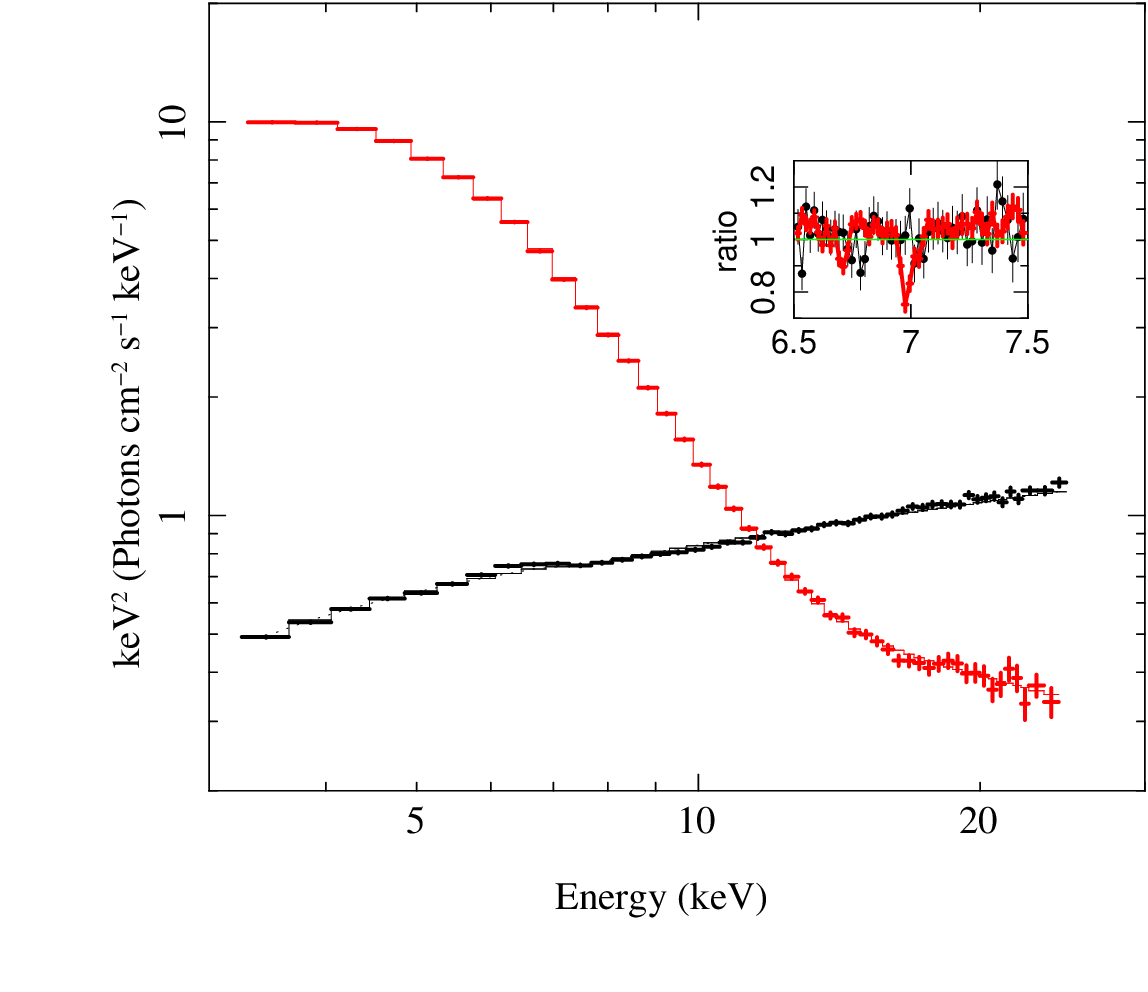} \\
\end{tabular}
\caption{The low/hard (black) and high/soft (red) spectra from RXTE
  data on H1743-322, with the inset showing the simultaneous Chandra
  data, with a clear difference in wind absorption features. The high
  energy 8.8-30~keV photo-ionising flux is quite similar between the
  two datasets, but the overall luminosity is quite different. If the
  wind stayed constant, responding only to the high energy
  photo-ionising flux, then the difference in wind properties would
  require a change in the wind structure, potentially linked to the
  appearance of the jet. However, thermal winds respond to the
  overall flux, as well as to the high energy part, and this predicts
  that the column should be smaller by a factor of $\sim 5$ in the
  low/hard state, consistent with the observations. }
\label{fig:h1743}
\end{figure}

We extract the simultaneous RXTE/Chandra data for these observations
in order to constrain the continuum shape and luminosity as well as
the wind features.  The standard pipeline RXTE continuum spectra are
shown in Fig.\ref{fig:h1743} (ObsIDs P95368-01-01-00 and
P80135-02-01-11 for the low/hard and high/soft states, respectively),
with the inset showing the TGCAT coadded HEG $\pm$~1 Chandra high
resolution spectra (ObsIDs 3803 and 11048, respectively) around the
iron line bandpass.  It is obvious that there is a large change in
bolometric luminosity as well as the state change and the change in
wind absorption features.  The wind should clearly have higher column
in this particular high/soft state than the comparison low/hard state
simply because the source has higher bolometric luminosity (see
Fig.\ref{fig:bhb_over}e and \ref{fig:bhb_obs}a).

Integrating the
model to get an unabsorbed bolometric flux gives $9.7\times
10^{-9}$~ergs~s$^{-1}$~cm$^{-2}$ 
for the low/hard 
state using the {\sc nthcomp} model with an assumed
electron temperature of 100~keV, while the {\sc diskbb} model for the high/soft state 
gives $4.3\times
10^{-8}$~ergs~s$^{-1}$~cm$^{-2}$. However, the 
intrinsic luminosity of the high/soft state is probably higher 
due to projection effects 
as the disc is seen at high inclination
- as evidenced by the fact it has a
disc wind (Ponti et al 2012) and a strong low frequency QPO (Ingram et
al 2016) as well as a high temperature disc (Munoz Darias et al
2013). The system parameters are not well known, but Dunn et al (2010)
show the hardness-intensity diagram for the outbursts look similar to
those of other BHB for a canonical 10\Msol mass and 5~kpc
distance. This then gives $L/L_{Edd}=0.02$ (low/hard) and $0.1$
(high/soft) without any projected area correction, or $L/L_{Edd}\sim
0.4$ with a cosine dependence assuming $i\sim 75^\circ$ (Steiner et al 2012).  This fits
rather well with the observed spectrum, as the {\sc diskbb}
temperature is 1.2~keV, i.e. $2\times$ higher than that assumed at the
transition at $L/L_{Edd}=0.02$, so the luminosity should be
$2^4=16\times$ higher, at $L/L_{Edd}\sim 0.3$.
The tail in the high/soft state carries roughly 5\% of the total bolometric power, so these
two spectra are very comparable to those assumed in the simulations shown in 
Figs.\ref{fig:bhb_over} and \ref{fig:bhb_obs}. 

The thermal wind model then predicts that there should be a wind column of
$\sim 8\times 10^{22}$~cm$^{-2}$ in the high/soft state, a factor 5
larger than the predicted column of $1.6\times 10^{22}$~cm$^{-2}$ in
the low/hard state. This is despite the fact that the high energy
luminosities are similar in the low/hard and high/soft states.
Thermal winds respond to changes in total luminosity and spectral
shape and are not just dependent on the high energy photo-ionising flux. The models
predict that the high energy photo-ionisation parameter is lower by a factor
5 in the high/soft state, as there is a stronger wind, but with
similar high energy flux.  However, this photo-ionisation is not
sufficient to fully strip the wind in the low/hard state, so the clear
prediction of the thermal wind model is that the absorption features
should be $\sim 5\times$ bigger in the high/soft state at $L/L_{Edd}=0.3-0.4$
compared to the brightest low/hard state at $L/L_{Edd}=0.02$, despite
them having similar high energy fluxes. This is easily compatible with
the Miller et al (2012) results.

\subsection{High luminosities and the wind in GRO J1655-40}

The dotted lines on Fig.\ref{fig:bhb_obs} show the effect of the
simple radiation pressure correction (see Section 4).  The column
increases dramatically as $L\to 0.7L_{Edd}$, becoming optically thick
to electron scattering, with a corresponding drop in ionisation
state. The dotted lines on Fig.\ref{fig:bhb_over} show that this much
stronger wind is launched from progressively smaller radii.

Our assumed disc dominated spectra at high luminosity are too simple
to describe those seen from GRS1915+105, so here we concentrate only
on the comparison to GRO J1655-40, which had a very soft spectrum at
the time when it produced the most extreme wind seen from this or any
other black hole binary. This anomolous wind has large column, $\log
N_h\sim 23.8$ and (compared to other binary winds) low ionisation
state, $\log \xi=4$ (Miller et al 2006; Miller et al 2008; Kallman et
al 2009). This ionisation state is comparable to those predicted here
for the simple radiation pressure correction at $L\to 0.7L_{Edd}$, but
the observed column is not optically thick, and the observed
luminosity is only $\sim 0.05 L_{Edd}$ (Miller et al 2006; Miller et
al 2008; Kallman et al 2009). 

Nonetheless, it seems possible that this could still be at least
partly (there are multiple velocity components: Kallman et al 2009;
Miller et al 2015; 2016) due to a thermal-radiative wind. The true
source luminosity will be underestimated if the wind become optically
thick and this optically thick material could be hidden by being
completely ionised, with the observed absorption lines arising from an
outer thin skin of partially ionised material. In this picture, an
optically thick wind is launched by thermal-radiative driving from the
inner disc, and expands outwards. Electron scattering in the
completely ionised, optically thick material reduces the observed
X-ray flux along the line of sight, with the observed absorption lines
arising in an outer, optically thin, photosphere of the wind. The
detection of the metastable FeXXII shows that this photosphere must
have $n\sim 10^{15}$~cm$^{-3}$ (Miller et al 2008; Kallman et al 2009)
and it is illuminated by the same (supressed) continuum which we see
as it is along the line of sight. Hence the observed $\log \xi=4$
implies a position for this photosphere at $R\sim 8\times 10^8$~cm
($\equiv 800R_g$), as in the magnetic wind models. The difference in this
thermal-radiative picture is that the wind is actually launched from even
closer to the black hole, and we see only the outer shell of the
expanding optically thick material.

This may seen an unnecessarily complex picture given the success of
the magnetic wind models in fitting the data quantatively
(e.g. Fukumura et al 2017). However, while the magnetic models can
quantatitively fit to the observed optically thin wind with the
observed low luminosity, they do not explain why this wind is only
seen in this one observation of GRO J1655-40.  There are other
similarly low luminosity Chandra datasets from this source which show
much higher ionisation winds, consistent with a thermal driving
(Nielsen \& Homan 2012), and similarly low luminosity Chandra data
from other sources which show only the expected thermal wind
signatures of H- and He-like iron (e.g. Ponti et al 2012). The
magnetic wind model also does not explain the other unusual features
of GRO J1655-40 in this dataset, namely the unusual lack of
variability seen in the corona of this state (Uttley \& Klein-Wolt
2015), or its unusually steep spectrum (Neilsen \& Homan 2012), or
give a geometry where the absorption lines only partially cover the
source (Kallman et al 2009). By contrast, all these other features can
be explained fairly naturally in an optically thick wind
model. Downscattering in the wind steepens the spectrum, supresses
variability, and makes an extended source geometry for partial
covering, and it is rare for BHB to reach (or exceed) Eddington which
explains why this wind has such different properties.

Nontheless, a thermal-radiative wind from a source with $L\to L_{Edd}$
remains a speculative explanation for this observation of GRO
J1655-40. To test whether this scenario can work quantatively requires
a hydrodynamic code to calculate the two dimensional wind structure,
including radiative continuum driving (e.g. Proga \& Kallman 2004)
with full Monte- Carlo radiation transport to handle the scattered
flux (e.g. Higgingbottom et al 2014). However, the limitations of the
magnetic wind models in explaining why the wind in this observation of GRO
J1655-40 is so different from other observations of this and other BHB's
motivate consideration of the possibility.

%==============================================
\section{Conclusions}
%==============================================

We use the analytic models for thermal winds of B83 and W96, combined
with a very simple geometric/kinematic model for the structure of the
wind to predict the column density, ionisation and velocity along any
line of sight at any luminosity for any spectrum.  We combine this
with a very simple model of the spectral evolution with luminosity in
BHB, including the major change in Compton temperature at the
hard-soft spectral transition, as well as the smaller but systematic
change in Compton temperature with luminosity within each state.

We show that the column density of the wind seen at any luminosity
generally increases with increasing mass accretion rate except for a
dip just after the transition to the high/soft state, where the much
lower Compton temperature suppresses the wind. This predicts that there
is more wind material just after the source makes a transition to
the low/hard state, in direct contrast to claims that the wind is
suppressed in the low/hard state and seen only in the high/soft state.
While photo-ionisation also plays a role in the visibility of the wind, 
we show that this is probably not enough to supress a wind just after the
transition to the low/hard state if it was visible in the high/soft state just 
beforehand. 

We critically examine the data on which the claims of wind suppression
are based.  GRS1915+105 (Neilsen \& Lee 2009) is close to Eddington,
and has complex spectra so the very simple models used here are
probably not applicable. Instead, H1743-322 (Miller et al 2012) shows
cannonical low/hard and high/soft states, with Chandra data clearly
showing that the wind in the high/soft state is absent in low/hard data
with similar high energy luminosity.  However, these two spectra are
very different in total luminosity, and this difference is enhanced by
the cosine dependence of the disc flux in the high/soft state, while
the low/hard state emission is more isotropic.  We estimate that these
two spectra differ by an order of magnitude in intrinsic luminosity,
so the much stronger wind in the high/soft state is entirely in line
with the thermal wind predictions. 

This removes any need for suppression of the wind at the transition
via magnetic fields switching from powering the wind in
the high/soft state, to powering the jet in the low/hard state
(Neilsen \& Lee 2009). Indeed, such a switch would be very surprising,
as the jet is almost certainly launched from the inner disc, while the
low outflow velocities seen in the wind shows that it is most likely
launched from the outer disc.

Radiation pressure should become important as the source approaches
the Eddington limit, increasing the mass loss rate as material above
the disc is unbound at progressively smaller radii.  We include a
simple correction for this which predicts stronger winds launched from
smaller radii as $L\to L_{Edd}$. This may be able to explain the
anomolous wind seen in GRO J1655-40 (Miller et al 2006) if the wind
becomes optically thick, supressing the observed luminosity (Shidatsu
et al 2016). However, a quantative comparison requires hydrodynamic
simulations in the Eddington regime which are beyond the scope of this
paper. Alternatively, Fukumura et al (2017) have shown quantatatively
that magnetic driving can be consistent with the broad properties of
this anomlous wind, though their model does not explain why similar
winds are not seen in this or other BHB at similar luminosities, nor
does it address the likelihood of the required magnetic field
configuration.

We conclude that there is at present no strong requirement for
magnetic winds in the majority of BHB, 
and it is possible that they are not required
in the supermassive black holes either (e.g Hagino et al 2015;
2016). Known wind launching mechanisms (thermal, UV line driven and
Eddington continuum) should be explored in detail before ruling them
out in favour of magnetic winds.

\section*{Acknowledgements}

We acknowledge
funding under Kakenhi 24105007, and 
CD acknowledges STFC funding under grant ST/L00075X/1 and a JSPS long
term fellowship L16581. We thank Daniel Proga and Norm Murray for very useful conversations, and 
especially thank Tod Woods for going above and beyond the call of duty
in answering detailed questions on the results of his paper.

%-----------------------------------------------------

\label{lastpage}
\end{document}